\documentclass[runningheads]{llncs}
\usepackage{orcidlink}
\usepackage[T1]{fontenc}
\usepackage{appendix}
\usepackage[table]{xcolor}
\usepackage{graphicx}
\usepackage{amsmath}
\usepackage{comment}
\usepackage{multirow}
\usepackage{hyperref}
\usepackage{url}
\usepackage[export]{adjustbox}
\usepackage{color,soul}
\usepackage{amsfonts}
\usepackage{amssymb}
\usepackage{graphicx}
\usepackage{float}
\usepackage{hyperref}
\usepackage{listings}
\usepackage{xcolor}
\usepackage{booktabs}
\usepackage{algorithm}
\usepackage{algpseudocode}
\DeclareMathOperator*{\argmax}{argmax}
 
\usepackage[inline,shortlabels]{enumitem}
\usepackage{xurl}

\usepackage{ntheorem}
\theoremseparator{:}

\definecolor[named]{mygreen}{rgb}{0.4,0.9,0.7}
\definecolor[named]{myred}{rgb}{1,0.7,0.5}

\begin{document}
%

\title{ \textbf{\normalfont\scshape AgriIR}: A Scalable Framework for Domain-Specific Knowledge Retrieval}

\author{
    Shuvam Banerji Seal$^\dagger$
    \inst{1}
    \orcidID{0009-0000-0714-569X}
    \and
    Aheli Poddar$^\dagger$
    \inst{2}
    \orcidID{0009-0009-6262-4103}
    \and
    Alok Mishra$^\dagger$
    \inst{1}
    \orcidID{0009-0002-7472-0812} 
    \and
    Dwaipayan Roy
    \inst{1}
    \orcidID{0000-0002-5962-5983}
}
\authorrunning{Banerji~Seal et al.}

\institute{
	Indian Institute of Science Education and Research, Kolkata, India \and
	Institute of Engineering \& Management, Kolkata, India\\
	\email{sbs22ms076@iiserkol.ac.in},
	\email{aheli.poddar2022@iem.edu.in},
    \email{maa24ms215@iiserkol.ac.in},
    \email{dwaipayan.roy@iiserkol.ac.in}\\
    \footnotetext{$^\dagger$ Contributed equally}
}

\maketitle              
\setcounter{footnote}{0}

\begin{abstract}
This paper introduces \textsc{AgriIR}, a configurable retrieval augmented generation (RAG) framework designed to deliver grounded, domain-specific answers while maintaining flexibility and low computational cost.
Instead of relying on large, monolithic models, \textsc{AgriIR} decomposes the information access process into declarative modular stages -- query refinement, sub-query planning, retrieval, synthesis, and evaluation.
This design allows practitioners to adapt the framework to new knowledge verticals without modifying the architecture.
Our reference implementation targets Indian agricultural information access, integrating 1B-parameter language models with adaptive retrievers and domain-aware agent catalogues. The system enforces deterministic citation, integrates telemetry for transparency, and includes automated deployment assets to ensure auditable, reproducible operation. By emphasizing architectural design and modular control, \textsc{AgriIR} demonstrates that well-engineered pipelines can achieve domain-accurate, trustworthy retrieval even under constrained resources. 
We argue that this approach exemplifies ``AI for Agriculture'' by promoting accessibility, sustainability, and accountability in retrieval-augmented generation systems.

\keywords{Large Language Model \and 
Domain-Specialized RAG \and
Multi-Agent Architecture \and
Intelligent Web Retrieval \and
Low-Resource Deployment \and
Information Retrieval for Good
.}
\end{abstract}


\section{Introduction}

Agriculture remains foundational to global livelihoods and food systems, employing approximately 916 million people worldwide in 2023 around 26.1\% of total global employment~\cite{fao2025employmentindicators}.
In India, which has the world's largest farming population, agriculture supports 58\% of the rural population and contributes roughly 18\% to the national GDP~\cite{goi2023agriculture}.
Despite this significance, the agricultural sector, particularly in low and middle-income countries, continues to struggle with information accessibility related to farming and support.
Farmers, extension officers, and policymakers often lack timely and contextually relevant knowledge for making decisions on crop management, irrigation, and climate resilience.
Bridging this information divide is thus not only a technological challenge but also a social imperative.

In recent years, large language models (LLMs) such as GPT-4 have shown remarkable performance across reasoning and knowledge-intensive tasks, offering new possibilities for intelligent agricultural advisory systems.
However, their direct application to real-world agricultural decision-making remains constrained by three systemic limitations.
First, \emph{resource requirements}: state-of-the-art models with 70B+ parameters require substantial computational infrastructure that is unavailable in rural contexts.
Second, \emph{domain drift}: general-purpose models lack specialized agricultural expertise, often yielding generic or misleading outputs.
Third, \emph{information reliability}: without explicit grounding in trustworthy sources, LLM hallucinations risk propagating erroneous recommendations that can have severe economic and environmental consequences~\cite{lin2025llmbasedagentssufferhallucinations}.

To address these limitations, the research community has increasingly turned toward retrieval-augmented generation (RAG), a hybrid paradigm that combines traditional Information Retrieval (IR) with neural generation to produce grounded and explainable responses~\cite{lewis2020retrievalaugmented,izacard2021leveraging}.
RAG systems offer a promising pathway toward domain-grounded knowledge access by retrieving authoritative documents and conditioning generation on retrieved evidence.
However, most existing RAG implementations remain tailored for high-resource environments, relying on large vector databases, extensive datasets, and GPU-heavy compute infrastructures~\cite{thakur2021beir}.
Consequently, their deployment in low-resource or domain-sensitive settings, such as agriculture, public health, or climate adaptation—remains impractical~\cite{koopman2023}.
This is precisely why applying LLMs for domain-specific applications in resource-constrained countries is particularly challenging, as such contexts amplify issues of computational scarcity, data scarcity and the need for locally valid knowledge grounding.

At the same time, there is growing recognition that architectural modularity and domain adaptability are critical to making RAG systems sustainable and reusable.
Existing frameworks like PyTerrier~\cite{macdonald2021pyterrier} and benchmarking efforts such as BEIR~\cite{thakur2021beir} have demonstrated the power of declarative experimentation in IR.
Building upon these insights, we introduce \textsc{AgriIR}, a configurable RAG framework that decomposes the end-to-end process into modular, declarative stages i.e., query refinement, retrieval, synthesis, and evaluation, where each controlled through configuration rather than hard-coded logic.
This design enables flexible reconfiguration across domains and datasets without re-engineering the system runtime.

Beyond technical modularity, the responsible IR community has emphasized the necessity of transparency, fairness, and efficiency in information systems~\cite{mitchell2019,bender2021,bernard2025}.
Large-scale generative architectures, however, remain opaque and energy-intensive, raising environmental and epistemic concerns~\cite{strubell2019energy,Bommasani2021FoundationModels,mai2024oppo}.
\textsc{AgriIR} directly responds to these concerns by adopting the principle of architectural intelligence over parameter scale.
Instead of scaling up parameters, \textsc{AgriIR} scales up design intelligence: it leverages systematic domain specialization, controlled sampling (temperature tuning), and multi-stage reasoning to enable 1B-parameter models to produce reliable, evidence-grounded responses comparable to much larger systems.

Through this approach, \textsc{AgriIR} operationalizes key tenets of responsible IR -- \emph{reproducibility}, \emph{accountability}, and \emph{reusability} -- by exposing every stage as a transparent, configurable interface~\cite{olteanu2021}.
While the framework can be generalized to multiple knowledge domains, this paper focuses on its deployment in Indian agriculture, a domain characterized by linguistic diversity, fragmented data ecosystems, and limited compute capacity.
We argue that such a context offers a uniquely stringent testbed for evaluating how architectural design, rather than parameter scale, can make RAG systems both socially relevant and technically sustainable.

\textsc{AgriIR} demonstrates that domain-specific retrieval systems can achieve reliability through architectural design rather than model scale. Instead of large general purpose models, \textsc{AgriIR} employs a lightweight RAG pipeline built on four principles:
\begin{itemize}
\item \textbf{Modular Task Decomposition}: Complex queries decompose into auditable stages (refinement, decomposition, retrieval, synthesis, evaluation) with stable interfaces enabling independent component substitution~\cite{adhikary2024iiserk}.
\item \textbf{Temperature Stratification}: Configurable temperature controls balance precision and creativity across stages, allowing behavioral tuning without code changes.
\item \textbf{Pluggable Retrieval}: A unified capability registry integrates vector databases, structured corpora, and live APIs, enabling automatic fallback and graceful degradation across heterogeneous sources.
\item \textbf{Declarative Domain Adaptation}: Domain artifacts -- prompts, agents, scoring heuristics—inject via configuration, enabling rapid vertical retargeting without fine-tuning or data collection.
\end{itemize}
These principles demonstrate that architectural intelligence compensates for parameter count in resource-constrained environments. By externalizing control through declarative configuration, \textsc{AgriIR} achieves competitive performance using 1B-parameter models against systems 7–70× larger. This design-driven approach enables practitioners to adapt the framework to new domains, i.e. agriculture, health, and climate, through reconfiguration alone. It prioritises interpretability, auditability and efficiency alongside accuracy.

Our codebase is publicly available to ensure full reproducibility and to support further research\footnote{\url{https://github.com/Shuvam-Banerji-Seal/AgriIR}}.
A video demonstration providing a walkthrough of the system is available here\footnote{\url{https://bit.ly/AgriIR}}.

\section{Related Work}
\subsection{Domain-Specific IR and RAG Systems}

Retrieval-augmented generation (RAG) has emerged as a dominant approach for grounding LLM outputs in external knowledge \cite{lewis2020retrievalaugmented}. While general-purpose RAG systems combine dense retrieval with large language models (typically 7B-70B parameters), they face deployment challenges in resource-constrained agricultural contexts. Domain-specific applications -- crop disease diagnosis \cite{mustofa2023comprehensivereviewplantleaf}, fertilizer recommendations \cite{katharria2025informationfusionsmartagriculture} demonstrate RAG's potential but typically rely on large models and static knowledge bases~\cite{peng2023embeddingbasedretrievalllmeffective}.

Recent agricultural LLM systems have explored specialized architectures for farming applications~\cite{KUSKA2024108924,Shaikh2025,Wilson2024}. ShizishanGPT \cite{yang2024shizishangptagriculturallargelanguage}, an agricultural large language model integrating tools and resources, implements a comprehensive modular RAG framework consisting of five key components: (1) a GPT-4-based module for handling general agricultural queries, (2) search engines that compensate for the limitations of static LLM knowledge by providing real-time updates, (3) agricultural knowledge graphs for structured domain facts, (4) retrieval modules using RAG to supplement domain knowledge, and (5) specialized agricultural agents that invoke domain-specific models for tasks such as crop phenotype prediction and gene expression analysis. The system was evaluated using a dataset of 100 agricultural questions, demonstrating superior performance over general-purpose LLMs through its integrated approach to domain knowledge and tool utilization. Similarly, AgroLLM \cite{samuel2025agrollmconnectingfarmersagricultural} develops an AI-powered agricultural chatbot using FAISS~\cite{douze2024faiss,johnson2019billion} vector databases and RAG, achieving 93\% accuracy with ChatGPT-4o Mini across four agricultural domains. These systems demonstrate the effectiveness of integrating multiple knowledge sources and specialized tools for agricultural applications.

Recent work emphasizes citation reliability \cite{survey_of_hallu}, with studies showing GPT-3.5/GPT-4 generate incorrect citations in 40-60\% of cases. \textsc{AgriIR} addresses this through post-hoc citation insertion using sentence similarity matching, operating independently of LLM generation.

\subsection{Model Efficiency and Architectural Design}

The LLM community increasingly recognizes that model size alone doesn't guarantee performance. Techniques like chain-of-thought prompting, temperature control for different reasoning stages, and specialized fine-tuning demonstrate how architectural choices improve smaller model performance.

Recent studies highlight the importance of energy efficiency in LLM deployment. Maliakel et al. \cite{maliakel2025investigatingenergyefficiencyperformance} investigate energy-performance trade-offs in LLM inference across different models (Falcon-7B\cite{almazrouei2023falconseriesopenlanguage}, Mistral-7B-v0.1\cite{jiang2023mistral7b}, LLaMA-3.2-1B \cite{grattafiori2024llama3herdmodels}, LLaMA-3.2-3B\cite{grattafiori2024llama3herdmodels}, GPT-Neo-2.7B) and tasks, analyzing input characteristics such as sequence length, entropy, and named entity density. Their work demonstrates that Dynamic Voltage and Frequency Scaling (DVFS) can reduce energy consumption by up to 30\% while preserving model quality, providing practical strategies for sustainable LLM inference.

\textsc{AgriIR} extends this philosophy to IR systems, showing that domain specialization, task decomposition, and temperature stratification enable 1B models to achieve comparable performance to naive deployments of 7B+ models in agricultural domains.

\subsection{Agricultural Information Systems}

Traditional agricultural knowledge systems -- ICAR repositories, mobile extension services, rule-based chatbots provide valuable content but lack natural language flexibility and real-time information integration \cite{icar2023}. Knowledge graphs enable semantic reasoning over agricultural relationships \cite{10.1007/978-3-030-18590-9_81} but require extensive manual curation.

Recent agricultural LLM systems have advanced beyond traditional approaches. ShizishanGPT \cite{yang2024shizishangptagriculturallargelanguage} and AgroLLM \cite{samuel2025agrollmconnectingfarmersagricultural} demonstrate the potential of specialized agricultural chatbots with RAG frameworks, achieving high accuracy through domain-specific knowledge integration. However, these systems typically rely on larger proprietary models and lack the configurable, multi-source retrieval architecture of \textsc{AgriIR}.
\textsc{AgriIR} bridges this gap by combining structured agricultural databases with real-time web information through an intelligent, resource-efficient architecture designed specifically for precise knowledge access.

In the subsequent section, we present a formal description of the proposed \textsc{AgriIR} framework. Building upon the motivation and challenges outlined earlier, this section delineates the system’s architectural design, operational workflow, and core algorithmic principles. We begin by describing the modular components that constitute the framework, including its retrieval pipeline, adaptive model selection strategy, and deterministic citation mechanism. This is followed by a detailed explanation of how \textsc{AgriIR} integrates large language models with structured agricultural databases and autonomous agents to ensure verifiable, domain-grounded information access. The formalization aims to provide a comprehensive understanding of the framework’s functionality, emphasizing its scalability, interpretability, and suitability for resource-constrained environments.

\section{\textbf{\normalfont\scshape AgriIR}: An IR System for Agricultural Knowledge Access}
\begin{figure}[h]
   \centering
   \includegraphics[width=\linewidth]{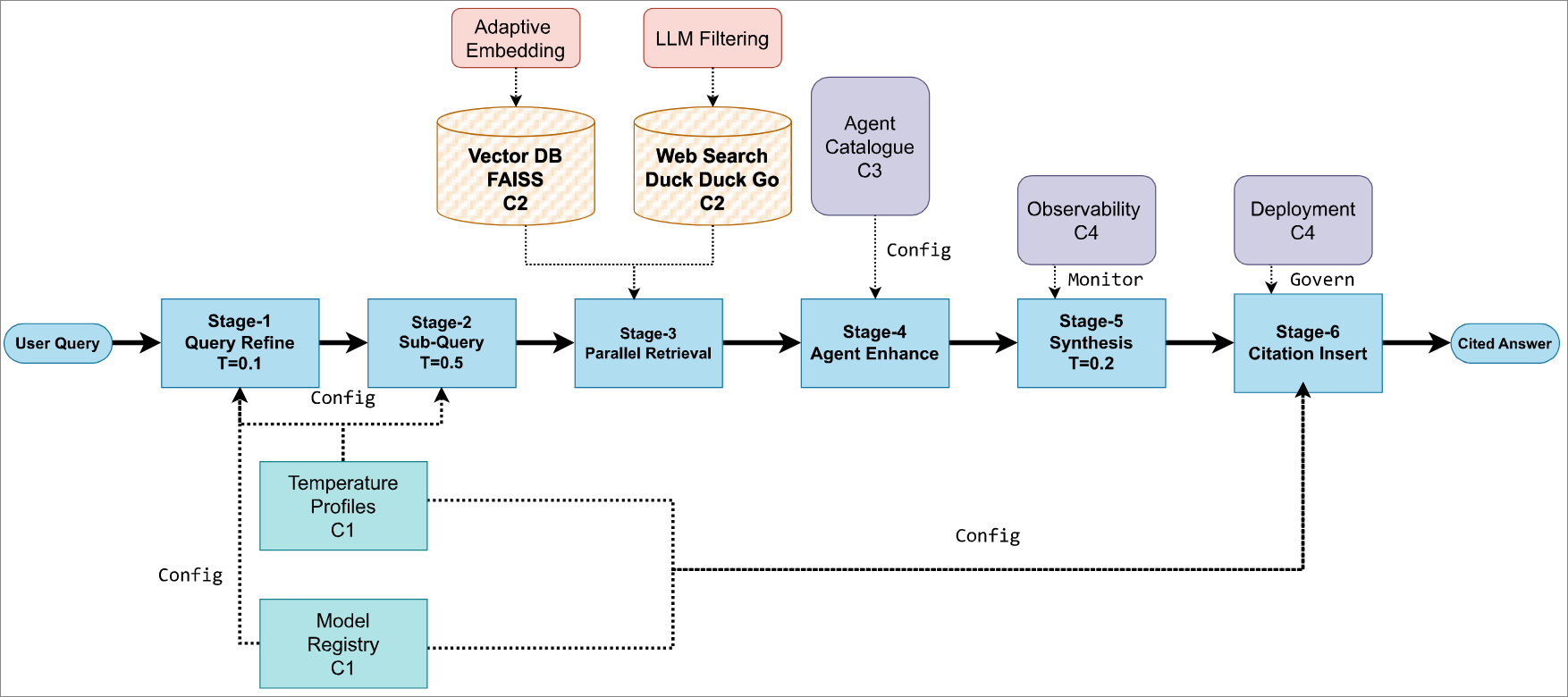}   
\caption{\small \textsc{AgriIR} Configurable Architecture Overview. All components are externally configurable without code modification.}
   \label{fig:architecture_overview}
\end{figure}

The complete system architecture of the proposed \textsc{AgriIR} framework is illustrated in Figure~\ref{fig:architecture_overview}.
To put the principles of modularity, configurability, and domain specialization into practice, we develop \textsc{AgriIR} for efficient use in resource-limited settings. \textsc{AgriIR} structures the entire retrieval and synthesis process as a declarative pipeline where each stage from query understanding to citation enforcement is controlled through configuration rather than code.

The system combines query decomposition, adaptive multi-source retrieval, domain-specialized agents, and lightweight generative models using Ollama\cite{ollama2023} framework to achieve superior performance without requiring large-scale parameters. Query refinement and decomposition break user inputs into tractable sub-problems; adaptive retrieval unifies structured databases and web sources through a pluggable registry; domain agents inject contextual expertise; and synthesis models generate responses with deterministic citation traceability. Temperature controls at each stage, enable precise behavioral tuning across refinement and synthesis.\\

\begin{algorithm}[H]
\caption{Complete \textsc{AgriIR} Pipeline}
\label{alg:agriir_pipeline}
\begin{algorithmic}[1]
\scriptsize
\Require Raw query $Q_{raw}$, $Q_{transliterated\_text} \gets Q_{voice}$, Models $M_{1b}$, $M_{27b}$, Agent registry $\mathcal{A}$, DB index $\mathcal{I}$
\Ensure Cited answer $A'$ with source index
\State \textit{// Stage 1: Query Refinement (Temperature = 0.1)}
\State $Q_{refined} \gets M_{1b}$\Call{Generate}{``Refine: " + $Q_{raw}$ or $Q_{transliterated\_text}$, temp=0.1}
\State
\State \textit{// Stage 2: Sub-Query Decomposition (Temperature = 0.5)}
\State $\{SQ_1, ..., SQ_n\} \gets M_{1b}$\Call{Generate}{``Decompose into 3-5 perspectives: " + $Q_{refined}$, temp=0.5}
\State
\State \textit{// Stage 3: Parallel Multi-Source Retrieval}
\For{$SQ_i$ \textbf{in parallel}}
\State \textit{// Database retrieval with adaptive embeddings}
\State $M_{embed} \gets$ \Call{SelectEmbeddingModel}{[models]} //higher MTEB ranking models go first based on hardware
\State $\mathcal{D}_i \gets$ \Call{FAISS\_Search}{$\mathcal{I}$, $M_{embed}$.encode($SQ_i$), k=3}
\State
\State \textit{// Web retrieval with intelligent filtering}
\State $candidates \gets$ \Call{Search\_Engine}{$SQ_i$ + ``agriculture site:.gov.in"}
\State $selected \gets M_{1b}$\Call{SelectArticles}{$candidates$, $SQ_i$, top=5}
\State $\mathcal{W}_i \gets$ \Call{ExtractContent}{$selected$}
\EndFor
\State $\mathcal{D} \gets \bigcup_i \mathcal{D}_i$, $\mathcal{W} \gets \bigcup_i \mathcal{W}_i$
\State
\State \textit{// Stage 4: Domain Agent Enhancement}
\For{$SQ_i \in \{SQ_1, ..., SQ_n\}$}
\State $agent^* \gets \argmax_{agent \in \mathcal{A}}$ \Call{KeywordScore}{$agent$, $SQ_i$}
\State $SQ_i \gets SQ_i$ + $agent^*$.domain\_keywords
\EndFor
\State
\State \textit{// Stage 5: Answer Synthesis (Temperature = 0.2)}
\State $M_{synth} \gets$ \Call{SelectModel}{$Q_{refined}$, [$M_{1b}$, $M_{27b}$]}
\State $A \gets M_{synth}$\Call{Generate}{``Synthesize from: " + $\mathcal{D}$ + $\mathcal{W}$, temp=0.2}
\State
\State \textit{// Stage 6: Deterministic Citation Insertion}
\State $encoder \gets$ SentenceTransformer(``transformer\_model")
\For{$sent \in$ \Call{SplitSentences}{$A$}}
\State $matches \gets \{s \in \mathcal{D} \cup \mathcal{W} \mid$ \Call{CosineSim}{$encoder$($sent$), $encoder$($s$)} $> 0.75\}$
\If{$matches \neq \emptyset$}
    \State Insert citation IDs from $matches$ after $sent$
\EndIf
\EndFor
\State
\Return Cited answer $A'$ with citation index
\end{algorithmic}
\end{algorithm}

Algorithm~\ref{alg:agriir_pipeline} formally defines the \textsc{AgriIR} workflow, which transforms an input query into a grounded, cited, and verifiable response through six sequential stages.
Each stage operates under a defined temperature regime and employs modular components that can be independently configured, substituted, or extended without altering the overall runtime architecture.
The subsequent discussion elaborates the core mechanisms underpinning this workflow, including temperature stratification, parallelization strategy, agentic data curation, adaptive retrieval, and deterministic citation.

\paragraph{\textbf{Stage 1: Query Refinement.}}
The process begins with a natural language user query \( Q_{\text{raw}} \) or \( Q_{\text{voice}} \)\footnote{transliterated to english using \url{https://docs.sarvam.ai/api-reference-docs/introduction}}, reflecting a real-world agricultural information need (e.g., ``\emph{How can smallholder farmers reduce nitrogen fertilizer usage in rice cultivation?}'').  
A lightweight 1B-parameter model refines this query under a deterministic temperature of 0.1, producing a clearer and more structured form \( Q_{\text{refined}} \).  
This refinement removes ambiguity and clarifies intent for subsequent sub-query decomposition and retrieval.  
The step is intentionally low-temperature to enforce consistency and avoid creative drift.

\paragraph{\textbf{Stage 2: Sub-Query Decomposition.}}
Next, the refined query is decomposed into multiple sub-queries \( \{SQ_1, SQ_2, \ldots, SQ_n\} \), each representing a distinct aspect of the original question.  
This step, executed by the same 1B model but with a moderate temperature of 0.5, introduces controlled diversity.  
For example, a single agricultural query may be split into sub-topics such as soil management, pest control, irrigation, and policy support.  
Decomposition allows \textsc{AgriIR} to handle multi-faceted queries by distributing retrieval across complementary information scopes.

\paragraph{\textbf{Stage 3: Parallel Multi-Source Retrieval.}}
Each sub-query is processed in parallel to minimize latency.  
For every \( SQ_i \), the system performs retrieval from both structured databases and web sources.

First, \textsc{AgriIR} selects the most appropriate embedding model from a ranked list of embedding models, depending on hardware availability.  
The chosen model encodes each sub-query into a dense representation, which is then searched against the FAISS index \( \mathcal{I} \).  
The top-\(k\) (default = 3) passages are returned, along with metadata such as source URL, authority score, and similarity value.

In parallel, a web retriever issues domain-constrained searches (e.g., ``site:.gov.in agriculture'') using \texttt{DuckDuckGo API}\footnote{\href{https://pypi.org/project/duckduckgo-search/}{https://pypi.org/project/duckduckgo-search/}}. 
Candidate URLs are filtered through an LLM-based selector that ranks them by relevance, credibility, and contextual fit.  
The top five articles are downloaded and parsed using \texttt{BeautifulSoup}\footnote{\href{https://pypi.org/project/beautifulsoup4/}{https://pypi.org/project/beautifulsoup4/}} and PDF extraction tools, yielding processed documents \( \mathcal{W}_i \).

All retrieved database and web passages are unified across sub-queries as \( \mathcal{D} = \bigcup_i \mathcal{D}_i \) and \( \mathcal{W} = \bigcup_i \mathcal{W}_i \).  
Parallelization via a \texttt{ThreadPoolExecutor} reduces total retrieval latency from $\sim$180\,s (sequential) to $\sim$50\,s for a four-subquery workload in our testing.

\paragraph{\textbf{Stage 4: Domain-Agent Enhancement.}}
To infuse domain expertise, \textsc{AgriIR} employs a registry of specialized agents \( \mathcal{A} \), each representing a knowledge domain (e.g., crop specialist, soil expert, pest manager, sustainability advisor, etc.).  
For each \( SQ_i \), the system computes a keyword-overlap score with the agents’ domain vocabularies.  
The agent with the highest score, \( agent^* \), is selected, and its contextual keywords are appended to the sub-query retrieval scope.  
This mechanism enables domain-aware query expansion without retraining the model.

\paragraph{\textbf{Stage 5: Answer Synthesis.}}
All retrieved materials \( \mathcal{D} \) and \( \mathcal{W} \) are passed to the synthesis module, which generates the final answer.  
\textsc{AgriIR} automatically selects between two generation models depending on the parameters: a 1B model for technical or factual queries, and a higher parameterized model (> 1B) for policy or contextual questions.  
Synthesis runs at a temperature of 0.2, balancing factual precision and readability.  
The model is prompted to generate an 800$-$1200 word response that integrates evidence from all retrieved documents.

\paragraph{\textbf{Stage 6: Deterministic Citation Insertion.}}
Finally, \textsc{AgriIR} enforces deterministic citation tracking to prevent hallucination and ensure auditability.  
The generated answer \( A \) is segmented into sentences, each embedded using the SentenceTransformer~\cite{reimers-2019-sentence-bert,reimers-2020-multilingual-sentence-bert}.  
For every sentence, cosine similarity is computed with all source chunks in \( \mathcal{D} \cup \mathcal{W} \).
If similarity exceeds 0.75 (consistent with prior semantic similarity work~\cite{ValdiviaCabrera2025} and supported by pilot sensitivity analysis on held-out agricultural queries), the corresponding source IDs (e.g., [DB$_{ij}$], [WEB$_{ij}$] where $i$ and $j$ indicates respectively document and chunk id) are appended inline.  
Multiple citations are added when a sentence synthesizes several sources.  
The final output $A'$ is thus a verifiable and citation-backed answer that includes explicit source indices and associated metadata (URLs, publication dates, relevance scores).

\begin{figure}[h]
\centering
\includegraphics[width=1.0\textwidth]{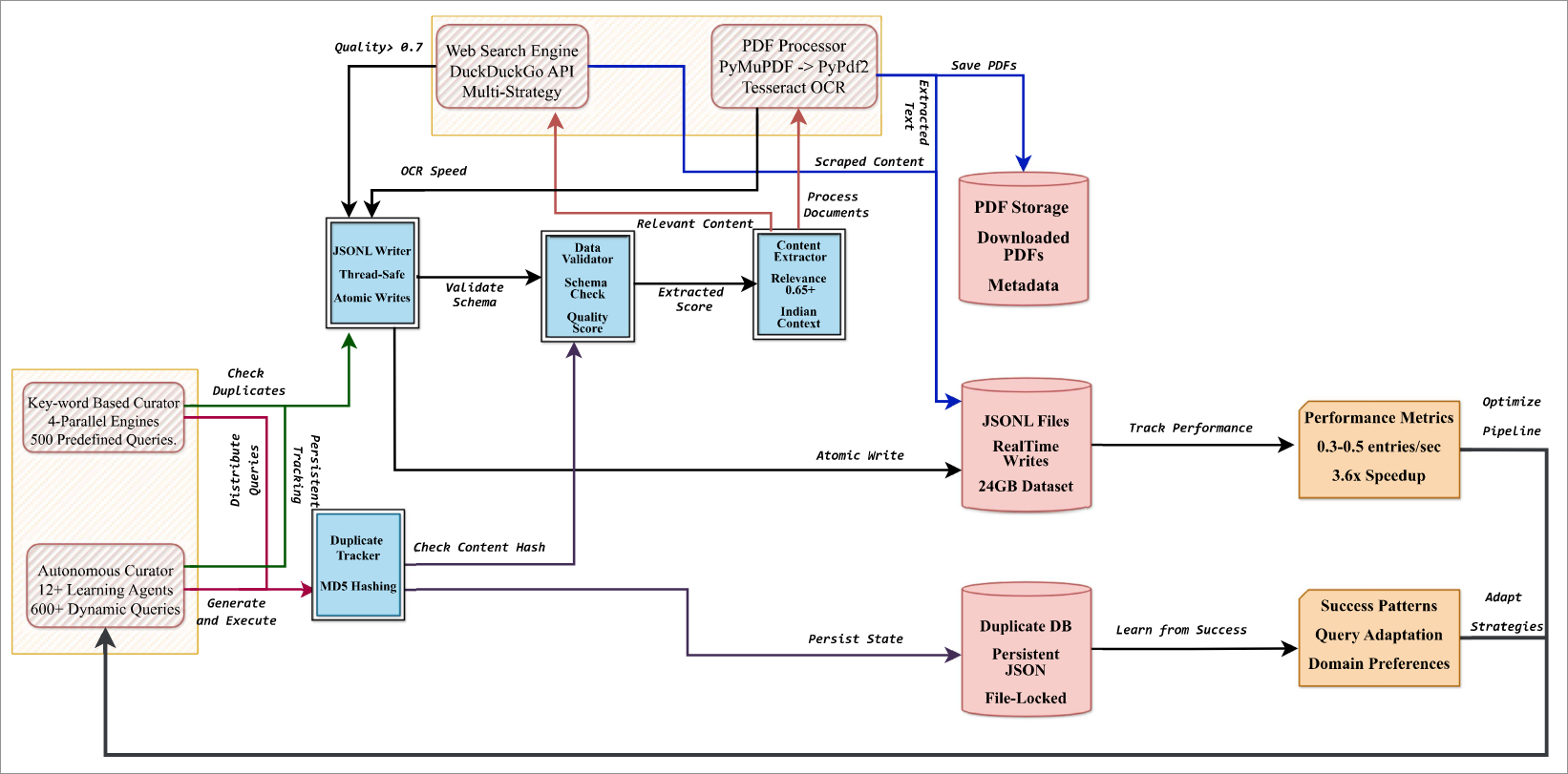}
\caption{Agentic Database Creation Architecture. Autonomous agents (purple) learn from success patterns via persistent tracking (red dashed line), while shared infrastructure (orange) ensures data quality and deduplication across both keyword-based and autonomous approaches.}
\label{fig:agentic_architecture}
\end{figure}

\paragraph{\textbf{Agentic Database Creation Architecture}} Next, we discuss the autonomous curation system, a multi-agent framework featuring persistent duplicate tracking, real-time JSONL logging, and quality-driven learning. Figure~\ref{fig:agentic_architecture} illustrates the complete pipeline.


Following are the key architectural features:

\begin{enumerate}
\item \textbf{Persistent Duplicate Tracking:} Cross-run deduplication using MD5 content hashing\cite{Rivest:1992:MMA:RFC1321} with file-locked JSON persistence. Prevents redundant processing across restarts with 4-method detection: URL normalization, URL hash, content hash, and title matching.

\item \textbf{Atomic JSONL Writing:} Thread-safe realtime persistence with fsync guarantees. Each entry undergoes duplicate checking before atomic write with file locking, ensuring data integrity under parallel agent execution.

\item \textbf{Multi-Library PDF Processing:} Hierarchical fallback between different OCR libraries with configurable OCR limit. Handles both text-extractable and image-based PDFs, storing processed content with metadata enrichment.

\item \textbf{Quality-Based Learning:} Autonomous agents track success patterns, failure patterns, and domain preferences. Quality scoring (0.0-1.0) considers content length (20\%), agriculture relevance (30\%), Indian context (20\%), data richness (20\%), and PDF bonus (10\%). Agents adapt future queries based on learned patterns.

\item \textbf{Graceful Degradation:} Adaptive embedding model selection based on MTEB leaderboard\cite{mteb_leaderboard} with automatic GPU detection. System remains operational even when preferred models are unavailable.
\end{enumerate}

\section{Results}
\subsection{Benchmark Dataset and Annotation}

We curated 191 agricultural queries through a multi-source approach: (1) SERP API\footnote{\url{https://serpapi.com/}} and Reddit API\footnote{\url{https://www.reddit.com/dev/api/}} for community-driven context; (2) 20 Indian agriculture associated government websites for precise official policy queries; (3) scraped 400+ candidate response, articles etc. across agricultural domains; (4) manually annotated top 191 queries from the responses, ensuring diversity across MSP policies, agricultural reforms, export regulations, climate adaptation, economic impacts, and institutional factors\footnote{\url{https://github.com/XAheli/AgriIR_Query_Gen}}.

For the annotation, we have considered a graded score (non-binary).
30 Annotators were selected from undergraduate programs in agriculture, food science, and pharmacy. Each question-answer pair from the pool of 191 was assigned to three annotators with relevant domain background based on its theme. Each annotator independently scored: (i) Answer satisfaction (0-4: worst to best) i.e. assessed based on relevance to the query, factual correctness, clarity, and completeness; (ii) Citation satisfaction (0-2: worst to best) i.e. reflected the appropriateness and usefulness of cited sources for supporting the answer. Reported scores represent the mean across all 30 annotators, providing robust human evaluation proxy. Inter-annotator agreement was measured using Cohen’s kappa~    \cite{doi:10.1177/001316446002000104}, and disagreements were resolved through discussion. Reported results reflect the mean scores across annotations.  Furthermore, the evaluation emphasized user-centric response quality and citation usefulness, rather than expert-level policy adjudication, making the annotation setup appropriate for the intended real-world use case.


\subsection{Evaluated Systems and Metrics}

We evaluate \textsc{AgriIR} across three complementary dimensions: answer quality, citation quality, and system efficiency. Together, these measures provide a holistic understanding of both the informational reliability and operational practicality of the framework in real-world agricultural contexts. \\
The evaluation compares multiple configurations of the \textsc{AgriIR} framework against both open and commercial baselines. We assess variants using \textsc{Llama3.2:3B}~\cite{grattafiori2024llama3herdmodels} (with and without database integration), \textsc{Gemma3:1B}~\cite{gemmateam2025gemma3technicalreport} (without database), and \textsc{Gemma3:27B} (with database). These are contrasted with three standalone large-model baselines: \textsc{ChatGPT-4o}, \textsc{Gemini~2.5~Flash}, and \textsc{GPT-OSS-120B}. This setup enables a systematic examination of how model scale and retrieval augmentation jointly influence performance under constrained resources.


Evaluation is based on a composite performance score that integrates both answer and citation quality, capturing factual accuracy and grounding relevance. The final score is computed based on a linear combination as in Equation~\ref{eq:finalscore}.

\begin{equation}
    \label{eq:finalscore}
  \text{Score} = \lambda \times \texttt{Answer} + (1-\lambda) \times \texttt{Citation} 
\end{equation}
The value of $\lambda$ was set to $0.7$ following preliminary experimentation on a held-out subset, chosen to balance factual correctness with citation grounding while avoiding overfitting to any particular evaluation split.
Statistical significance of observed differences is tested using standard inferential methods, including the Student’s $t$-test, Welch’s $t$-test, and the Mann–Whitney $U$ test, with effect sizes reported using Cohen’s~$d$. This ensures robust comparison across model types and deployment conditions.

\subsection{Performance Results}
The comprehensive result is presented in Table~\ref{tab:main_results}. From the table, it can be observed that \textsc{AgriIR}\_Gemma-3-27B achieves a composite score of $0.820\pm0.208$, which is statistically equivalent to ChatGPT-4o ($0.840\pm0.233$, $p=0.493$), while significantly outperforming Gemini 1.5 Flash ($0.779\pm0.250$) and GPT-OSS-120B ($0.705\pm0.246$, $p<0.001$). Critically, \textsc{AgriIR} models excel in citation quality ($73$-$84\%$ perfect citations), a capability absent in baseline models. Table~\ref{tab:significance} validates \textsc{AgriIR}'s architectural efficiency: Gemma-3-27B matches ChatGPT-4o performance ($p=0.493$, $d=0.08$) and significantly exceeds the $120$B-parameter GPT-OSS baseline ($p<0.001$, $d=0.45$). Within the framework, scaling from $1$B to $27$B parameters yields substantial gains ($\Delta=+0.172$, $p<0.001$, $d=0.82$), though database integration shows negligible impact for Llama3.2:3B ($p=0.204$, $d=0.14$).

\begin{table}[H]
\centering
\caption{
Comprehensive evaluation of all systems over 191 queries annotated by 30 human evaluators. Reported metrics include mean and standard deviation for answer quality (0–4) and citation quality (0–2), along with the percentage of responses rated as good (>=3) and perfect (2), respectively. The composite score (0–1) is computed as $\lambda \times (\texttt{Answer}) + (1-\lambda) \times (\texttt{Citation})$; we report results with setting $\lambda = 0.7$. Baseline models do not support citation tracking and are therefore evaluated only on answer quality.
}
\label{tab:main_results}
\resizebox{\textwidth}{!}{
\begin{tabular}{@{}lcccccc@{}}
\toprule
\multirow{2}{*}{\textbf{Model}} & \multicolumn{2}{c}{\textbf{Answer (0-4)}} & \multicolumn{2}{c}{\textbf{Citation (0-2)}} & \multicolumn{2}{c}{\textbf{Composite (0-1)}} \\
\cmidrule(lr){2-3} \cmidrule(lr){4-5} \cmidrule(lr){6-7}
& Mean$\pm$Std & \% Good ($\geq$3) & Mean$\pm$Std & \% Perfect (2) & Mean$\pm$Std & Rank \\
\midrule
ChatGPT-4o & 3.36$\pm$0.94 & 88.5 & --- & --- & 0.840$\pm$0.233 & 1 \\
\textbf{\textsc{AgriIR}\_Gemma3:27B} & \textbf{3.24$\pm$0.87} & \textbf{86.9} & \textbf{1.69$\pm$0.55} & \textbf{73.0} & \textbf{0.820$\pm$0.208} & \textbf{2} \\
Gemini 2.5 Flash & 3.12$\pm$1.01 & 78.7 & --- & --- & 0.779$\pm$0.250 & 3 \\
\textsc{AgriIR}\_Llama3.2:3B (w/o DB) & 2.72$\pm$1.12 & 68.0 & 1.62$\pm$0.64 & 69.7 & 0.718$\pm$0.252 & 4 \\
GPT-OSS-120B & 2.82$\pm$0.99 & 70.5 & --- & --- & 0.705$\pm$0.246 & 5 \\
\textsc{AgriIR}\_Llama3.2:3B (w/ DB) & 2.62$\pm$1.09 & 63.9 & 1.51$\pm$0.68 & 59.0 & 0.684$\pm$0.258 & 6 \\
\textsc{AgriIR}\_Gemma3:1B (w/o DB) & 2.44$\pm$1.05 & 49.2 & 1.47$\pm$0.65 & 54.9 & 0.648$\pm$0.233 & 7 \\
\bottomrule
\end{tabular}
}
\end{table}


\begin{table}[H]
\centering
\caption{
Statistical significance analysis for pairwise comparisons among evaluated systems. 
The table reports mean score differences ($\Delta$~Mean), corresponding $p$-values from Student’s \textit{t}-test, Welch’s test, and the Mann–Whitney~\textit{U} test, along with effect sizes computed using Cohen’s~$d$ \cite{GESSESSE201981}. 
Significance levels are denoted as *$p{<}0.05$, **$p{<}0.01$, and ***$p{<}0.001$. 
Effect size interpretation follows conventional thresholds: negligible~($d{<}0.2$), small~($0.2{\leq}d{<}0.5$), medium~($0.5{\leq}d{<}0.8$), and large~($d{\geq}0.8$).
}
\label{tab:significance}
\resizebox{\textwidth}{!}{
\begin{tabular}{@{}lccccc@{}}
\toprule
\textbf{Comparison} & \textbf{Δ Mean} & \textbf{t-test} & \textbf{Welch} & \textbf{Mann-W} & \textbf{Cohen's d} \\
\midrule
\multicolumn{6}{c}{\textit{\textsc{AgriIR} vs Baselines}} \\  \cmidrule{1-6}
Gemma3:27B vs ChatGPT-4o & -0.020 & 0.493 & 0.493 & 0.381 & 0.08 (negl.) \\
Gemma3:27B vs Gemini 2.5 & +0.041 & 0.046* & 0.046* & 0.038* & 0.16 (small) \\
Gemma3:27B vs GPT-OSS & +0.115 & <0.001*** & <0.001*** & <0.001*** & 0.45 (med.) \\
\hline
\hline
\multicolumn{6}{c}{\textit{Within \textsc{AgriIR} Framework}} \\ \cmidrule{1-6}
Gemma3:27B vs Gemma3:1B & +0.172 & <0.001*** & <0.001*** & <0.001*** & 0.82 (large) \\
Gemma3:27B vs Llama3.2:3B(w/ DB) & +0.136 & <0.001*** & <0.001*** & <0.001*** & 0.63 (med.) \\
Llama3.2:3B(w/ DB) vs (w/o DB) & -0.034 & 0.204 & 0.204 & 0.123 & 0.14 (negl.) \\
\bottomrule
\end{tabular}
}
\end{table}

\section{Conclusion and Future Work}
    This paper presented \textsc{AgriIR}, a domain-specialized information retrieval framework addressing critical challenges in agricultural information access. Our work makes three primary contributions to trustworthy information retrieval. First, we developed a deterministic citation mechanism using sentence-level semantic similarity that operates independently of LLM generation, achieving 59-73\% perfect citation accuracy and eliminating citation hallucination through direct measurement of semantic overlap between generated content and retrieved sources. Second, we demonstrated intelligent multi-phase web retrieval combining multi-strategy candidate gathering, LLM-based article selection, and comprehensive content extraction to reduce retrieval noise while prioritizing authoritative agricultural sources. Third, we showed scalable autonomous knowledge acquisition through specialized agents with persistent duplicate tracking, collecting 15,247 agricultural entries without manual curation to address knowledge staleness in static RAG systems.
    
    Evaluation on 191 agricultural policy queries with 30 human annotators validates our approach. \textsc{AgriIR}\_Gemma3:27B achieves statistical parity with ChatGPT-4o (composite scores: $0.820 \pm 0.208$ vs $0.840 \pm 0.233$, $p=0.493$) while significantly outperforming GPT-OSS-120B despite using $4.4×$ fewer parameters ($\Delta=0.115, p<0.001, d=0.45$). This demonstrates that domain specialization through multi-agent architectures and intelligent retrieval outperforms brute-force parameter scaling. Critically, \textsc{AgriIR} variants provide verifiable citations—a capability absent in baseline models—essential for trustworthy agricultural decision-making where information accuracy impacts farmer livelihoods.

Several extensions can further strengthen \textsc{AgriIR}. Multimodal integration could combine satellite imagery (Normalized Difference Vegetation Index~\cite{GESSESSE201981}, soil moisture), IoT sensor data, and visual question answering for crop disease diagnosis~\cite{Upadhyay2025}, supported by joint cross-modal embeddings~\cite{Salvador2017-rl} for unified retrieval and citation. 
While the domain agent enhancement mechanism is effective, it requires continuous maintenance.
Further, the domain-specific keyword lists and the agent configurations must be regularly updated; otherwise, outdated knowledge or incomplete coverage can introduce inconsistencies and degrade the retrieval performance over time.
Causal reasoning through structural causal models would enable policy counterfactuals such as ``\emph{How would a 10\% MSP reduction affect wheat yields?}'' Federated learning across agricultural universities could enhance query understanding while preserving privacy through secure computation. Citation graph analytics could reveal source reliability patterns, guiding autonomous data collection. Personalization by region, crop type, or farm size would allow tailored recommendations for diverse user groups.

Beyond agriculture, \textsc{AgriIR}'s principles i.e. deterministic citation, domain grounding, and autonomous data acquisition - apply to safety-critical fields like healthcare, law, and education, where trust and verifiability are essential. The framework and benchmark of 191 annotated policy queries advance the vision of ``Information Retrieval for Good,'' showing that domain intelligence and verifiable architecture can rival large general-purpose models while ensuring reliability, efficiency, and real-world usability.

\subsubsection*{\textbf{Acknowledgement.}}
We gratefully acknowledge the Computation and Data Science Department at IISER Kolkata for providing the computational resources necessary to carry out the experiments reported in this work. We also thank Abhinav Dhingra for illustrating Figures \ref{fig:agentic_architecture} and \ref{fig:architecture_overview}.

\subsubsection*{\textbf{Disclosure of Interests.}}
The authors declare no competing interests that influenced the research, authorship, or publication of this article.

\bibliographystyle{splncs04}
\bibliography{main}

@misc{gemmateam2025gemma3technicalreport,
      title={Gemma 3 Technical Report}, 
      author={Gemma Team},
      year={2025},
      eprint={2503.19786},
      archivePrefix={arXiv},
      primaryClass={cs.CL},
      url={https://arxiv.org/abs/2503.19786}, 
}

@techreport{goi2023agriculture,
  title       = {Annual Report 2022-23},
  author      = {{Department of Animal Husbandry and Dairying}},
  institution = {Ministry of Fisheries, Animal Husbandry and Dairying, Government of India},
  year        = {2023},
  url         = {https://dahd.gov.in/sites/default/files/2023-06/FINALREPORT2023ENGLISH.pdf}
}

@misc{fao2025employmentindicators,
  author       = {{Food and Agriculture Organization of the United Nations}},
  title        = {Employment Indicators 2000–2023 (July 2025 Update)},
  year         = {2025},
  howpublished = {FAOSTAT Highlights Archive},
  institution  = {Food and Agriculture Organization of the United Nations},
  url          = {https://www.fao.org/statistics/highlights-archive/highlights-detail/employment-indicators-2000-2023-%28july-2025-update%29/en},
  note         = {Accessed: 2025-10-27}
}

@misc{icar2023,
  title        = {Indian Council of Agricultural Research},
  author       = {{ICAR}},
  year         = {2023},
  howpublished = {Official Website},
  url          = {https://icar.org.in/},
  note         = {Accessed: 2025-10-27}
}

@inproceedings{adhikary2024iiserk,
  title={{IISERK@ToT\_2024}: Query Reformulation and Layered Retrieval for Tip-of-Tongue Items},
  author={Adhikary, Subinay and Banerji Seal, Shuvam and Sar, Soumyadeep and Roy, Dwaipayan},
  booktitle={Proceedings of the Thirty-Third Text REtrieval Conference (TREC 2024)},
  year={2024},
  organization={National Institute of Standards and Technology (NIST)},
  url = {https://trec.nist.gov/pubs/trec33/papers/IISER-K.tot.pdf}
}

@inproceedings{lewis2020retrievalaugmented,
 author = {Lewis, Patrick and Perez, Ethan and Piktus, Aleksandra and Petroni, Fabio and Karpukhin, Vladimir and Goyal, Naman and K\"{u}ttler, Heinrich and Lewis, Mike and Yih, Wen-tau and Rockt\"{a}schel, Tim and Riedel, Sebastian and Kiela, Douwe},
 booktitle = {Advances in Neural Information Processing Systems},
 editor = {H. Larochelle and M. Ranzato and R. Hadsell and M.F. Balcan and H. Lin},
 pages = {9459--9474},
 publisher = {Curran Associates, Inc.},
 title = {Retrieval-Augmented Generation for Knowledge-Intensive NLP Tasks},
 url = {https://proceedings.neurips.cc/paper_files/paper/2020/file/6b493230205f780e1bc26945df7481e5-Paper.pdf},
 volume = {33},
 year = {2020}
}

@article{survey_of_hallu,
author = {Ji, Ziwei and Lee, Nayeon and Frieske, Rita and Yu, Tiezheng and Su, Dan and Xu, Yan and Ishii, Etsuko and Bang, Ye Jin and Madotto, Andrea and Fung, Pascale},
title = {Survey of Hallucination in Natural Language Generation},
year = {2023},
issue_date = {December 2023},
publisher = {Association for Computing Machinery},
address = {New York, NY, USA},
volume = {55},
number = {12},
issn = {0360-0300},
url = {https://doi.org/10.1145/3571730},
doi = {10.1145/3571730},
journal = {ACM Comput. Surv.},
month = mar,
articleno = {248},
numpages = {38}
}

@InProceedings{10.1007/978-3-030-18590-9_81,
author="Chen, Yuanzhe
and Kuang, Jun
and Cheng, Dawei
and Zheng, Jianbin
and Gao, Ming
and Zhou, Aoying",
editor="Li, Guoliang
and Yang, Jun
and Gama, Joao
and Natwichai, Juggapong
and Tong, Yongxin",
title="AgriKG: An Agricultural Knowledge Graph and Its Applications",
booktitle="Database Systems for Advanced Applications",
year="2019",
publisher="Springer International Publishing",
address="Cham",
pages="533--537",
isbn="978-3-030-18590-9",
  doi       = {10.1007/978-3-030-18590-9_81},
  url       = {https://doi.org/10.1007/978-3-030-18590-9_81}
}

@misc{mustofa2023comprehensivereviewplantleaf,
      title={A comprehensive review on Plant Leaf Disease detection using Deep learning}, 
      author={Sumaya Mustofa and Md Mehedi Hasan Munna and Yousuf Rayhan Emon and Golam Rabbany and Md Taimur Ahad},
      year={2023},
      eprint={2308.14087},
      archivePrefix={arXiv},
      primaryClass={cs.CV},
      url={https://arxiv.org/abs/2308.14087}, 
}

@misc{katharria2025informationfusionsmartagriculture,
title = {Information fusion in smart agriculture: machine learning applications and future research directions},
journal = {Information Fusion},
volume = {129},
pages = {104040},
year = {2026},
issn = {1566-2535},
doi = {https://doi.org/10.1016/j.inffus.2025.104040},
url = {https://www.sciencedirect.com/science/article/pii/S1566253525011029},
author = {Aashu Katharria and Millie Pant and Juan D. Velásquez and Václav Snášel and Kanchan Rajwar and Kusum Deep},
}

@misc{samuel2025agrollmconnectingfarmersagricultural,
      title={AgroLLM: Connecting Farmers and Agricultural Practices through Large Language Models for Enhanced Knowledge Transfer and Practical Application}, 
      author={Dinesh Jackson Samuel and Inna Skarga-Bandurova and David Sikolia and Muhammad Awais},
      year={2025},
      eprint={2503.04788},
      archivePrefix={arXiv},
      primaryClass={cs.CL},
      url={https://arxiv.org/abs/2503.04788}, 
}

@misc{maliakel2025investigatingenergyefficiencyperformance,
      title={Investigating Energy Efficiency and Performance Trade-offs in LLM Inference Across Tasks and DVFS Settings}, 
      author={Paul Joe Maliakel and Shashikant Ilager and Ivona Brandic},
      year={2025},
      eprint={2501.08219},
      archivePrefix={arXiv},
      primaryClass={cs.LG},
      url={https://arxiv.org/abs/2501.08219}, 
}

@misc{yang2024shizishangptagriculturallargelanguage,
      title={ShizishanGPT: An Agricultural Large Language Model Integrating Tools and Resources}, 
author = {Yang, Shuting and Liu, Zehui and Mayer, Wolfgang and Ding, Ningpei and Wang, Ying and Huang, Yu and Wu, Pengfei and Li, Wanli and Li, Lin and Zhang, Hong-Yu and Feng, Zaiwen},
year = {2024},
isbn = {978-981-96-0572-9},
publisher = {Springer-Verlag},
address = {Berlin, Heidelberg},
url = {https://doi.org/10.1007/978-981-96-0573-6_21},
doi = {10.1007/978-981-96-0573-6_21},
booktitle = {Web Information Systems Engineering – WISE 2024: 25th International Conference, Doha, Qatar, December 2–5, 2024, Proceedings, Part IV},
pages = {284–298},
numpages = {15},
location = {Doha, Qatar}
}

@article{KUSKA2024108924,
title = {AI for crop production – Where can large language models (LLMs) provide substantial value?},
journal = {Computers and Electronics in Agriculture},
volume = {221},
pages = {108924},
year = {2024},
issn = {0168-1699},
doi = {https://doi.org/10.1016/j.compag.2024.108924},
url = {https://www.sciencedirect.com/science/article/pii/S0168169924003156},
author = {Matheus Thomas Kuska and Mirwaes Wahabzada and Stefan Paulus}
}

@article{Shaikh2025,
  author       = {Tawseef Ayoub Shaikh and Tabasum Rasool and K. Veningston and Syed Mufassir Yaseen},
  title        = {The role of large language models in agriculture: harvesting the future with LLM intelligence},
  journal      = {Progress in Artificial Intelligence},
  year         = {2025},
  volume       = {14},
  number       = {2},
  pages        = {117--164},
  doi          = {10.1007/s13748-024-00359-4},
  url          = {https://doi.org/10.1007/s13748-024-00359-4},
  issn         = {2192-6360}
}

@inproceedings{Wilson2024,
  author       = {Shyama Wilson and Athula Ginige and Jeevani Goonatilake},
  title        = {Design Science Research Approach for Ontology Development in Agriculture: Utilising advances of LLM for Automated Entity Extraction},
  booktitle    = {Proceedings of the Australasian Conference on Information Systems (ACIS 2024)},
  year         = {2024},
  number       = {150},
  publisher    = {Association for Information Systems},
  url          = {https://aisel.aisnet.org/acis2024/150},
  note         = {ACIS 2024 Proceedings, Paper 150}
}

@misc{peng2023embeddingbasedretrievalllmeffective,
      title={Embedding-based Retrieval with LLM for Effective Agriculture Information Extracting from Unstructured Data}, 
      author={Ruoling Peng and Kang Liu and Po Yang and Zhipeng Yuan and Shunbao Li},
      year={2023},
      eprint={2308.03107},
      archivePrefix={arXiv},
      primaryClass={cs.AI},
      url={https://arxiv.org/abs/2308.03107}, 
}

@inproceedings{izacard2021leveraging,
    title = "Leveraging Passage Retrieval with Generative Models for Open Domain Question Answering",
    author = "Izacard, Gautier  and
      Grave, Edouard",
    editor = "Merlo, Paola  and
      Tiedemann, Jorg  and
      Tsarfaty, Reut",
    booktitle = "Proceedings of the 16th Conference of the European Chapter of the Association for Computational Linguistics: Main Volume",
    month = apr,
    year = "2021",
    address = "Online",
    publisher = "Association for Computational Linguistics",
    url = "https://aclanthology.org/2021.eacl-main.74/",
    doi = "10.18653/v1/2021.eacl-main.74",
    pages = "874--880",
}

@article{koopman2023,
  title = {AgAsk: an agent to help answer farmer’s questions from scientific documents},
  volume = {25},
  ISSN = {1432-1300},
  url = {http://dx.doi.org/10.1007/s00799-023-00369-y},
  DOI = {10.1007/s00799-023-00369-y},
  number = {4},
  journal = {International Journal on Digital Libraries},
  publisher = {Springer Science and Business Media LLC},
  author = {Koopman,  Bevan and Mourad,  Ahmed and Li,  Hang and Vegt,  Anton van der and Zhuang,  Shengyao and Gibson,  Simon and Dang,  Yash and Lawrence,  David and Zuccon,  Guido},
  year = {2023},
  month = jun,
  pages = {569–584}
}

@inproceedings{macdonald2021pyterrier,
author = {Macdonald, Craig and Tonellotto, Nicola and MacAvaney, Sean and Ounis, Iadh},
title = {PyTerrier: Declarative Experimentation in Python from BM25 to Dense Retrieval},
year = {2021},
isbn = {9781450384469},
publisher = {Association for Computing Machinery},
address = {New York, NY, USA},
url = {https://doi.org/10.1145/3459637.3482013},
doi = {10.1145/3459637.3482013},
booktitle = {Proceedings of the 30th ACM International Conference on Information \& Knowledge Management},
pages = {4526–4533},
numpages = {8},
location = {Virtual Event, Queensland, Australia},
series = {CIKM '21}
}

@inproceedings{thakur2021beir,
title={{BEIR}: A Heterogeneous Benchmark for Zero-shot Evaluation of Information Retrieval Models},
author={Nandan Thakur and Nils Reimers and Andreas R{\"u}ckl{\'e} and Abhishek Srivastava and Iryna Gurevych},
booktitle={Thirty-fifth Conference on Neural Information Processing Systems Datasets and Benchmarks Track (Round 2)},
year={2021},
url={https://openreview.net/forum?id=wCu6T5xFjeJ}
}

@inproceedings{mitchell2019,
author = {Mitchell, Margaret and Wu, Simone and Zaldivar, Andrew and Barnes, Parker and Vasserman, Lucy and Hutchinson, Ben and Spitzer, Elena and Raji, Inioluwa Deborah and Gebru, Timnit},
title = {Model Cards for Model Reporting},
year = {2019},
isbn = {9781450361255},
publisher = {Association for Computing Machinery},
address = {New York, NY, USA},
url = {https://doi.org/10.1145/3287560.3287596},
doi = {10.1145/3287560.3287596},
booktitle = {Proceedings of the Conference on Fairness, Accountability, and Transparency},
pages = {220–229},
numpages = {10},
location = {Atlanta, GA, USA},
series = {FAT* '19}
}

@inproceedings{bender2021,
author = {Bender, Emily M. and Gebru, Timnit and McMillan-Major, Angelina and Shmitchell, Shmargaret},
title = {On the Dangers of Stochastic Parrots: Can Language Models Be Too Big? ��},
year = {2021},
isbn = {9781450383097},
publisher = {Association for Computing Machinery},
address = {New York, NY, USA},
url = {https://doi.org/10.1145/3442188.3445922},
doi = {10.1145/3442188.3445922},
booktitle = {Proceedings of the 2021 ACM Conference on Fairness, Accountability, and Transparency},
pages = {610–623},
numpages = {14},
location = {Virtual Event, Canada},
series = {FAccT '21}
}

@article{bernard2025,
author = {Bernard, Nolwenn and Balog, Krisztian},
title = {A Systematic Review of Fairness, Accountability, Transparency, and Ethics in Information Retrieval},
year = {2025},
issue_date = {June 2025},
publisher = {Association for Computing Machinery},
address = {New York, NY, USA},
volume = {57},
number = {6},
issn = {0360-0300},
url = {https://doi.org/10.1145/3637211},
doi = {10.1145/3637211},
journal = {ACM Comput. Surv.},
month = feb,
articleno = {136},
numpages = {29}
}

@inproceedings{strubell2019energy,
    title = "Energy and Policy Considerations for Deep Learning in {NLP}",
    author = "Strubell, Emma  and
      Ganesh, Ananya  and
      McCallum, Andrew",
    editor = "Korhonen, Anna  and
      Traum, David  and
      M{\`a}rquez, Llu{\'i}s",
    booktitle = "Proceedings of the 57th Annual Meeting of the Association for Computational Linguistics",
    month = jul,
    year = "2019",
    address = "Florence, Italy",
    publisher = "Association for Computational Linguistics",
    url = "https://aclanthology.org/P19-1355/",
    doi = "10.18653/v1/P19-1355",
    pages = "3645--3650"
}

@article{mai2024oppo,
author = {Mai, Gengchen and Huang, Weiming and Sun, Jin and Song, Suhang and Mishra, Deepak and Liu, Ninghao and Gao, Song and Liu, Tianming and Cong, Gao and Hu, Yingjie and Cundy, Chris and Li, Ziyuan and Zhu, Rui and Lao, Ni},
title = {On the Opportunities and Challenges of Foundation Models for GeoAI (Vision Paper)},
year = {2024},
issue_date = {June 2024},
publisher = {Association for Computing Machinery},
address = {New York, NY, USA},
volume = {10},
number = {2},
issn = {2374-0353},
url = {https://doi.org/10.1145/3653070},
doi = {10.1145/3653070},
journal = {ACM Trans. Spatial Algorithms Syst.},
month = jul,
articleno = {11},
numpages = {46}
}

@article{Bommasani2021FoundationModels,
title={On the Opportunities and Risks of Foundation Models},
author={Rishi Bommasani and Drew A. Hudson et al.},
journal={ArXiv},
year={2021},
url={https://crfm.stanford.edu/assets/report.pdf}
}

@article{olteanu2021,
author = {Olteanu, Alexandra and Garcia-Gathright, Jean and de Rijke, Maarten and Ekstrand, Michael D. and Roegiest, Adam and Lipani, Aldo and Beutel, Alex and Olteanu, Alexandra and Lucic, Ana and Stoica, Ana-Andreea and Das, Anubrata and Biega, Asia and Voorn, Bart and Hauff, Claudia and Spina, Damiano and Lewis, David and Oard, Douglas W. and Yilmaz, Emine and Hasibi, Faegheh and Kazai, Gabriella and McDonald, Graham and Haned, Hinda and Ounis, Iadh and van der Linden, Ilse and Garcia-Gathright, Jean and Baan, Joris and Lau, Kamuela N. and Balog, Krisztian and de Rijke, Maarten and Sayed, Mahmoud and Panteli, Maria and Sanderson, Mark and Lease, Matthew and Ekstrand, Michael D. and Lahoti, Preethi and Kamishima, Toshihiro},
title = {FACTS-IR: fairness, accountability, confidentiality, transparency, and safety in information retrieval},
year = {2021},
issue_date = {December 2019},
publisher = {Association for Computing Machinery},
address = {New York, NY, USA},
volume = {53},
number = {2},
issn = {0163-5840},
url = {https://doi.org/10.1145/3458553.3458556},
doi = {10.1145/3458553.3458556},
journal = {SIGIR Forum},
month = mar,
pages = {20–43},
numpages = {24}
}

@misc{lin2025llmbasedagentssufferhallucinations,
      title={LLM-based Agents Suffer from Hallucinations: A Survey of Taxonomy, Methods, and Directions}, 
      author={Xixun Lin and Yucheng Ning and Jingwen Zhang and Yan Dong and Yilong Liu and Yongxuan Wu and Xiaohua Qi and Nan Sun and Yanmin Shang and Pengfei Cao and Lixin Zou and Xu Chen and Chuan Zhou and Jia Wu and Shirui Pan and Bin Wang and Yanan Cao and Kai Chen and Songlin Hu and Li Guo},
      year={2025},
      eprint={2509.18970},
      archivePrefix={arXiv},
      primaryClass={cs.AI},
      url={https://arxiv.org/abs/2509.18970}, 
}

@misc{grattafiori2024llama3herdmodels,
      title={The Llama 3 Herd of Models}, 
      author={Aaron Grattafiori and Abhimanyu Dubey et al.},
      year={2024},
      eprint={2407.21783},
      archivePrefix={arXiv},
      primaryClass={cs.AI},
      url={https://arxiv.org/abs/2407.21783}, 
}

@misc{almazrouei2023falconseriesopenlanguage,
      title={The Falcon Series of Open Language Models}, 
      author={Ebtesam Almazrouei and Hamza Alobeidli and Abdulaziz Alshamsi and Alessandro Cappelli and Ruxandra Cojocaru and Mérouane Debbah and Étienne Goffinet and Daniel Hesslow and Julien Launay and Quentin Malartic and Daniele Mazzotta and Badreddine Noune and Baptiste Pannier and Guilherme Penedo},
      year={2023},
      eprint={2311.16867},
      archivePrefix={arXiv},
      primaryClass={cs.CL},
      url={https://arxiv.org/abs/2311.16867}, 
}

@misc{jiang2023mistral7b,
      title={Mistral 7B}, 
      author={Albert Q. Jiang and Alexandre Sablayrolles and Arthur Mensch and Chris Bamford and Devendra Singh Chaplot and Diego de las Casas and Florian Bressand and Gianna Lengyel and Guillaume Lample and Lucile Saulnier and Lélio Renard Lavaud and Marie-Anne Lachaux and Pierre Stock and Teven Le Scao and Thibaut Lavril and Thomas Wang and Timothée Lacroix and William El Sayed},
      year={2023},
      eprint={2310.06825},
      archivePrefix={arXiv},
      primaryClass={cs.CL},
      url={https://arxiv.org/abs/2310.06825}, 
}

@online{mteb_leaderboard,
  author    = {{Hugging Face}},
  title     = {Massive Text Embedding Benchmark (MTEB) Leaderboard},
  year      = {2025},
  url       = {https://huggingface.co/spaces/mteb/leaderboard},
  urldate   = {2025-10-27},
  note      = {Accessed on 27 October 2025}
}

@software{ollama2023,
  author       = {Ollama Team},
  title        = {Ollama: An Open Source Framework for Running and Serving Large Language Models Locally},
  year         = {2023},
  url          = {https://github.com/ollama/ollama},
  note         = {Version latest, accessed 27 October 2025},
}

@incollection{GESSESSE201981,
title = {Chapter 8 - Temporal relationships between time series CHIRPS-rainfall estimation and eMODIS-NDVI satellite images in Amhara Region, Ethiopia},
editor = {Assefa M. Melesse and Wossenu Abtew and Gabriel Senay},
booktitle = {Extreme Hydrology and Climate Variability},
publisher = {Elsevier},
pages = {81-92},
year = {2019},
isbn = {978-0-12-815998-9},
doi = {https://doi.org/10.1016/B978-0-12-815998-9.00008-7},
url = {https://www.sciencedirect.com/science/article/pii/B9780128159989000087},
author = {Agenagnew A. Gessesse and Assefa M. Melesse},
}

@inproceedings{reimers-2019-sentence-bert,
    title = "Sentence-BERT: Sentence Embeddings using Siamese BERT-Networks",
    author = "Reimers, Nils  and
      Gurevych, Iryna",
    editor = "Inui, Kentaro  and
      Jiang, Jing  and
      Ng, Vincent  and
      Wan, Xiaojun",
    booktitle = "Proceedings of the 2019 Conference on Empirical Methods in Natural Language Processing and the 9th International Joint Conference on Natural Language Processing (EMNLP-IJCNLP)",
    month = nov,
    year = "2019",
    address = "Hong Kong, China",
    publisher = "Association for Computational Linguistics",
    url = "https://aclanthology.org/D19-1410/",
    doi = "10.18653/v1/D19-1410",
    pages = "3982--3992",
}

@inproceedings{reimers-2020-multilingual-sentence-bert,
    title = "Making Monolingual Sentence Embeddings Multilingual using Knowledge Distillation",
    author = "Reimers, Nils and Gurevych, Iryna",
    booktitle = "Proceedings of the 2020 Conference on Empirical Methods in Natural Language Processing (EMNLP)",
    month = nov,
    year = "2020",
    address = "Online",
    publisher = "Association for Computational Linguistics",
    url = "https://aclanthology.org/2020.emnlp-main.365/",
    doi = "10.18653/v1/2020.emnlp-main.365",
    pages = "4512--4525"
}

@article{Rivest:1992:MMA:RFC1321,
  address = {United States},
  author = {Rivest, R.},
  number = {RFC1321},
  publisher = {RFC Editor},
  title = {The MD5 Message-Digest Algorithm},
  url = {http://www.ietf.org/rfc/rfc1321.txt},
  year = 1992
}

@article{douze2024faiss,
      title={The Faiss library},
      author={Matthijs Douze and Alexandr Guzhva and Chengqi Deng and Jeff Johnson and Gergely Szilvasy and Pierre-Emmanuel Mazaré and Maria Lomeli and Lucas Hosseini and Hervé Jégou},
      year={2024},
      eprint={2401.08281},
      archivePrefix={arXiv},
      primaryClass={cs.LG}
}

@article{johnson2019billion,
  title={Billion-scale similarity search with {GPUs}},
  author={Johnson, Jeff and Douze, Matthijs and J{\'e}gou, Herv{\'e}},
  journal={IEEE Transactions on Big Data},
  volume={7},
  number={3},
  pages={535--547},
  year={2019},
  publisher={IEEE}
}

@article{ValdiviaCabrera2025,
  author    = {Melissa Valdivia Cabrera and Michael Johnstone and Joshua Hayward and Kristy A. Bolton and Douglas Creighton},
  title     = {Integration of large-scale community-developed causal loop diagrams: a Natural Language Processing approach to merging factors based on semantic similarity},
  journal   = {BMC Public Health},
  year      = {2025},
  volume    = {25},
  number    = {1},
  pages     = {923},
  doi       = {10.1186/s12889-025-22142-3},
  url       = {https://doi.org/10.1186/s12889-025-22142-3},
  issn      = {1471-2458}
}

@article{Upadhyay2025,
  author    = {Upadhyay, Abhishek and Chandel, Narendra Singh and Singh, Krishna Pratap and Chakraborty, Subir Kumar and Nandede, Balaji M. and Kumar, Mohit and Subeesh, A. and Upendar, Konga and Salem, Ali and Elbeltagi, Ahmed},
  title     = {Deep learning and computer vision in plant disease detection: a comprehensive review of techniques, models, and trends in precision agriculture},
  journal   = {Artificial Intelligence Review},
  year      = {2025},
  volume    = {58},
  number    = {3},
  pages     = {92},
  doi       = {10.1007/s10462-024-11100-x},
  url       = {https://doi.org/10.1007/s10462-024-11100-x},
  issn      = {1573-7462}
}

@INPROCEEDINGS{Salvador2017-rl,
  title           = "Learning cross-modal embeddings for cooking recipes and
                     food images",
  booktitle       = "2017 {IEEE} Conference on Computer Vision and Pattern
                     Recognition ({CVPR})",
  author          = "Salvador, Amaia and Hynes, Nicholas and Aytar, Yusuf and
                     Marin, Javier and Ofli, Ferda and Weber, Ingmar and
                     Torralba, Antonio",
  publisher       = "IEEE",
  month           =  jul,
  year            =  2017,
  conference      = "2017 IEEE Conference on Computer Vision and Pattern
                     Recognition (CVPR)",
  location        = "Honolulu, HI"
}

@article{doi:10.1177/001316446002000104,
author = {Jacob Cohen},
title ={A Coefficient of Agreement for Nominal Scales},

journal = {Educational and Psychological Measurement},
volume = {20},
number = {1},
pages = {37-46},
year = {1960},
doi = {10.1177/001316446002000104},

URL = { 
    
        https://doi.org/10.1177/001316446002000104
    
    

},
eprint = { 
    
        https://doi.org/10.1177/001316446002000104
    
    

}

}
\end{document}